\documentstyle [12pt] {article}
\def\nicebreak{\pagebreak}
\begin{document}

{\obeylines
\hfill CLNS 93/1204
\hfill IP-ASTP-15-93
\hfill May, 1993}
\vskip 0.5 cm

\centerline{\large {\bf Heavy Quark and Chiral Symmetry Predictions}}
\medskip
\centerline{\large {\bf for Semileptonic Decays}
$\bar{B} \to D (D^\ast) \pi \ell \bar{\nu}$}

\bigskip\bigskip
\bigskip

\centerline{\bf Hai-Yang Cheng$^{a,d}$, Chi-Yee Cheung$^a$, William
Dimm$^c$, Guey-Lin Lin$^a$,}
\centerline{\bf Y. C. Lin$^b$, Tung-Mow Yan$^{a,c}$, and Hoi-Lai Yu$^a$}

\bigskip

\centerline{$^a$ Institute of Physics, Academia Sinica, Taipei,}
\centerline{Taiwan 11529, Republic of China}

\medskip

\centerline{$^b$ Physics Department, National Central University,
Chung-li,}
\centerline{Taiwan 32054, Republic of China}

\medskip

\centerline{$^c$ Floyd R. Newman Laboratory of Nuclear Studies, Cornell
University}
\centerline{Ithaca, New York 14853, USA}

\medskip

\centerline{$^d$ Institute for Theoretical Physics, State University of New
York}
\centerline{Stony Brook, New York 11794, USA}

\bigskip\bigskip
\bigskip\bigskip

\centerline{\bf Abstract}

\begin{quotation}

We study in detail the prediction for the semileptonic decays $\bar{B}
\to D (D^\ast) \pi \ell \bar{\nu}$ by heavy quark and chiral symmetry.
The branching ratio for $\bar{B} \to D \pi \ell \bar{\nu}$ is quite
significant, as big as $(0.5-1)\%$. The
branching ratio for $\bar{B} \to D^\ast \pi
\ell \bar{\nu}$ is only of order $10^{-4}-10^{-5}$.
Numerical results for various single particle spectra and their dependence
on the pion momentum cutoff schemes are presented in a
series of figures, as are the model independent
ratios for differential rates of $D$ and $D^\ast$.  We also study the
parity-violation effects on the decay rates for different polarization
states of the $D^\ast$.

\end{quotation}

\pagebreak

\noindent{\bf I.~~Introduction}

\bigskip

Flavor and spin symmetry of heavy quarks [1,2] and chiral symmetry of light
quarks together impose strong restrictions on the semileptonic decays of a
$\bar{B}$ meson [3-6] such as $\bar{B} \to D \pi\ell\bar{\nu}$ and $\bar{B} \to
D^\ast \pi\ell\bar{\nu}$.  Heavy quark symmetry predicts that the weak
vertex contains only a universal Isgur-Wise function with an overall
factor calculable in QCD.  Chiral symmetry, on the other hand, requires
one unknown coupling constant to describe soft pion emissions from
any ground-state heavy meson at low energies.  Thus, the semileptonic decays
with a soft pion are completely determined by the Isgur-Wise function
measured in $\bar{B}\to D^\ast \ell \bar{\nu}$ and the coupling constant that
describes the strong decay $D^\ast \to D \pi$.  The Feynman diagrams
for the decays to be studied in this work are shown in Figs. 1 and 2.
The matrix elements
for these decays are explicitly given in [3].  In this paper, we explore
the implications in detail.  Furthermore, since the unknowns appear as an
overall factor of the decay matrix elements, many of the ratios of the
differential spectra are free of any adjustable parameter.  Some of these
ratios are presented.

There is another motivation to study the semileptonic decays of the $\bar{B}$
meson with emission of additional pions.  The 1992 Particle Data Group (PDG)
[7] gives
$$
B (B^0 \to D^- \ell^+ \nu) = (1.8 \pm 0.5)\% ~,
\eqno(1.1)
$$
$$
B (B^0 \to D^{\ast -} \ell^+ \nu) = (4.9 \pm 0.8)\% ~,
\eqno(1.2)
$$
$$B (B^+\to \bar{D}^0\ell^+\nu)= (1.6\pm 0.7)\%~,\eqno(1.3)$$
$$B (B^+\to \bar{D}^{*0}\ell^+\nu)=(4.6\pm 1.0)\%~,\eqno(1.4)$$
$$
B (B \to e^\pm \nu_e ~{\rm hadrons})~ = (10.7 \pm 0.5)\% ~,
\eqno(1.5)
$$
\noindent where $\ell$ indicates $e$ or $\mu$ mode (not sum over modes), and
the charge of $B$ is not determined in the last branching
ratio (1.5). Clearly, beside the $D \ell \bar{\nu}$ and $D^\ast \ell \bar{\nu}$
modes, there exist other important semileptonic decays of the $\bar{B}$
meson.  There are indications from ARGUS [8] and CLEO [9] that $\bar{B}
\to D^{\ast \ast} \ell\bar{\nu}$ gives a significant contribution.
It is still interesting to ask how large the branching ratios for $\bar{B}
\to D \pi\ell\bar{\nu}$ and $\bar{B} \to D^\ast\pi\ell\bar{\nu}$
are.  It
turns out that the branching ratio for $\bar{B}\to D\pi\ell\bar{\nu}$
can be quite
significant, perhaps as large as $1\%$, while the branching ratio for
$\bar{B}\to
D^\ast\pi\ell\bar{\nu}$ is much smaller, of order $10^{-4}$ to $10^{-5}$.

   We would like to comment on our work in relation to other recent studies
on similar subjects [10,11,12]. Lee, Lu and Wise [10] have extended the
earlier formalisms for the $K_{\ell 4}$ and $D_{\ell 4}$ decays [13] to
$D\to K\pi\bar{\ell}\nu$, $D\to \pi\pi\bar{\ell}\nu$, $B\to\pi\pi\bar{\ell}
\nu$, and $\bar{B}\to D\pi\ell\bar{
\nu}$. For the last decay, Lee {\it et al.} [10] have considered the
predictions of chiral perturbation theory and heavy quark symmetry. They have
calculated the decay rate in the region where chiral perturbation theory is
expected to be valid. Kramer and Palmer [12] have discussed the decay
$\bar{B}\to D^\ast\ell\bar{\nu}\to D\pi\ell\bar{\nu}$ in the resonant region
and calculated the rate and angular correlation parameters in the framework
of chiral and heavy quark symmetry. Lee [11] has generalized the analysis of
[10] to the decay $\bar{B}\to D^\ast\pi\ell\bar{\nu}$ and has considered only
the decay rates with the $D^\ast$ polarizations summed over. In our
work, starting with the amplitudes given in [3], we derive explicit
formulae for the differential decay rates of both
$\bar{B}\to D\pi\ell\bar{\nu}$ and
$\bar{B}\to D^\ast\pi\ell\bar{\nu}$. Single particle energy spectra for $D$ or
$D^\ast$, the electron and the pion are evaluated numerically; their
dependence on the pion momentum cutoff schemes is studied. In the case
of $\bar{B}\to D\pi\ell\bar{\nu}$, the $D^\ast$ pole dominates the amplitude,
and the
rates for the $D\pi$ system in the resonant and nonresonant regions (to be
defined in Section V) are sensitive to the total decay width of $D^\ast$.
Although the charged $D^{*\pm}$ decay almost exclusively to $D\pi$, the
neutral $D^{*0}$ has a substantial radiative decay contribution [14,15].
Consequently,
the widths of $D^*$ are not simply related to the $D^*D\pi$ coupling constant.
With our theoretical results for the $D^*D\pi$ coupling constant and the
total widths of $D^{*\pm}$ and $D^{*0}$ [14], we are able to predict the
decay rates for $\bar{B}\to D^*\ell\bar{\nu}$ and $\bar{B}\to (D\pi)_{\rm
nonres}\ell\bar{\nu}$. The definitions of resonanting $D\pi$
(to be identified
with $D^*$) and nonresonanting $D\pi$ are given in Section V. Our results for
$\bar{B}\to D^*\ell\bar{\nu}$ for both charged and neutral $\bar{B}$ mesons
agree
with the available data (1.2) and (1.4). Our approach differs from those of
[10] and [12]. In [10] it is suggested that the decay rates for
$\bar{B}\to D\pi\ell\bar{\nu}$ can be used to fix the $D^* D\pi$ coupling
constant. This is possible in principle provided that the $D\pi$ system is
sufficiently far away from the $D^*$ resonance so that its rate is calculable
without knowing the $D^*$ widths. But then the rates are rather small. On the
other hand, the authors of [12] proposed that the decays $\bar{B}\to
(D\pi)_{\rm res}\ell\bar{\nu}$ can be used to determine the $D^*D\pi$ coupling
constant. As we have pointed out earlier, this will require information on
other decay modes of $D^*$.

   It is not straightforward to compare our numerical results with those
presented in [10,11,12].  As already mentioned, Lee, Lu and Wise [10] restrict
themselves to the kinematic region away from the $D^*$ pole in
$\bar{B} \to D\pi+e\bar{\nu}$.  However, we include the region close to the
resonance in
our calculation.  Lee [11] does not work out the total rates for
$\bar{B} \to D^*\pi+e\bar{\nu}$, and none of our single particle spectra
correspond to his tabulated differential rates.  A crude estimate indicates
that our total rate for $\bar{B} \to D^*\pi+e\bar{\nu}$ would have agreed
with his within about a factor of two if the same Isgur-Wise function
and value of the $D^*D\pi$ coupling constant were used.  Kramer and
Palmer [12] did not specify which charged modes they considered.  Our rates
for $\bar{B} \to D\pi+e\bar{\nu}$ in the resonant $D\pi$ region agree
with theirs within about a factor of two.

   In the case of $\bar{B}\to D^\ast\pi\ell\bar{\nu}$, the single particle
spectra are calculated for polarized $D^\ast$. The
underlying $V-A$ interaction of the quarks makes the
spectra polarization dependent. Instead of describing all
these results in words, we choose to present them in a series of figures.

The paper is organized as follows. In Section II we review the general
kinematics of the four-body semileptonic decays.  In Section III we make use
of the matrix elements given in [3] to calculate the differential decay
rates for $\bar{B} \to D \pi \ell\bar{\nu}$.  The same analysis is carried
out for $\bar{B} \to D^\ast \pi \ell\bar{\nu}$ in Section IV, with special
attention given to the polarization of $D^\ast$.  The results are discussed
in Section V. We present the single particle spectra, the semileptonic rates
versus the $D^*$ width, etc., in a series of
figures. The connection between different Lorentz frames is considered in
Appendix A. In Appendix B we derive a linear relation between the
semileptonic decay rate and the inverse of $D^*$ width and discuss some of its
implications. Some preliminary results have been reported earlier by two of
us [16].

\nicebreak

\noindent{\bf II.~~Kinematics}

\bigskip

In this section we will review the kinematics of the decays
$$
\bar{B} \to D + \pi + \ell \bar{\nu} ~,
\eqno(2.1a)
$$
$$
\bar{B} \to D^\ast + \pi + \ell \bar{\nu} ~.
\eqno(2.1b)
$$
\noindent General kinematics for such processes has been
studied by several authors before [10-13, 17]. We will pay special attention to
the new features for the latter where the polarization of $D^\ast$ is
involved.  It is well known that there are five independent kinematic
variables for these processes if the spin of the initial state is zero or
is not observed.
Let the momentum of the $\bar{B}$ meson, $D$ (or $D^\ast$), the
pion, the charged lepton, and the neutrino be $P_B, ~p, ~q,~p_\ell$ and
$p_\nu$, respectively.
For the five variables we follow earlier authors' convention and pick
\begin{description}
\item[$s_M = (p+q)^2$] ~~,
\item[$s_L = (p_\ell + p_\nu)^2$] ~~,
\item[$\theta$ = ] the angle between $\vec{p}$ in the $D (D^\ast) \pi$ rest
frame and the line of the flight of the $D(D^\ast) \pi$ system in the $\bar{B}$
meson's rest frame,
\item[$\theta_\ell$ =] the angle between $\vec{p_\ell}$ in the $\ell\bar{\nu}$
rest frame and the line of flight of the $\ell \bar{\nu}$ system in the
$\bar{B}$ meson's rest frame,
\item[$\phi$ =] the angle between the normals to the planes defined in the
$\bar{B}$ meson's rest frame by the momenta of the $D (D^\ast) \pi$ pair and
the $\ell\bar{\nu}$ pair, respectively.  The sense of $\phi$ is from the
$D(D^\ast) \pi$ plane to the $\ell \bar{\nu}$ plane.
\end{description}

These variables are depicted in Fig. 3.  In the same figure, we also display
the three orthonormal vectors associated with the 3-momentum of $D^\ast$ in the
$D^\ast \pi$ rest-frame.  They will be useful for describing the
polarization states of $D^\ast$.  The positive $z$-axis is along the line of
flight of the $D(D^\ast)
\pi$ system in the $\bar{B}$ meson's rest frame; the $x$-axis is in
the $D(D^\ast) \pi$ plane.  The lepton mass will be neglected.  We now form the
combinations,
$$
P = p + q ~, ~~ Q = p - q ~,
\eqno(2.2a)
$$
$$
L = p_\ell + p_\nu ~, ~~ N = p_\ell - p_\nu ~~,
\eqno(2.2b)
$$
\noindent and find
$$
P \cdot L = \frac{1}{2} (m^2_B -s_M - s_L )~,
\eqno(2.3a)
$$
$$
L \cdot N = 0 ~,
\eqno(2.3b)
$$
$$
P \cdot Q = m^2 - m^2_\pi ~,
\eqno(2.3c)
$$
$$
Q^2 = 2 (m^2 + m^2_\pi ) - s_M ~,
\eqno(2.3d)
$$
$$
N^2 = -s_L ~,
\eqno(2.3e)
$$
\noindent where $m^2 = m^2_D$ or $m^2_{D^\ast}$ as the case may be.

Three frames of reference are of particular interest:  the $B$-frame in
which the $\bar{B}$ meson is at rest, the $M$-frame which is the
center-of-mass frame of the $D(D^\ast)\pi$ system, and the $L$-frame which
is the center-of-mass frame
of the lepton pair.  To construct some of the Lorentz invariants, it is
often necessary to specify the individual components of various
four-vectors in one of these coordinate systems.  This information is
provided in Appendix A.  In particular, we find
$$
\begin{array}{ccl}
Q \cdot N & = & \left( \frac{m^2 - m^2_\pi}{s_M} \right) X \cos \theta_\ell +
\beta(P \cdot L) \cos \theta \cos \theta_\ell \\
 & & - \sqrt{s_M s_L} \beta \sin \theta \sin \theta_\ell \cos \phi ~,
\end{array}
\eqno(2.4a)
$$
$$
\sigma \equiv \epsilon_{\mu \nu \lambda \kappa} Q^\mu P^\nu N^\lambda
L^\kappa = - \sqrt{s_M s_L} \beta X \sin \theta \sin \theta_\ell \sin \phi
\eqno(2.4b)
$$
$$
P_B \cdot p  =   \frac{1}{2} \left[ \frac{s_M + m^2 - m^2_\pi}{2s_M}
\left( m^2_B + s_M - s_L \right) + X \beta \cos \theta \right]~,
\eqno(2.4c)
$$
$$
P_B \cdot q = \frac{1}{2} \left[ \frac{s_M + m^2_\pi - m^2}{2s_M} \left(
m^2_B + s_M - s_L \right) - X \beta \cos \theta \right] ~~,
\eqno(2.4d)
$$
$$
P_B \cdot p_\ell = \frac{1}{2} \left[ \frac{1}{2} \left( m^2_B + s_L - s_M
\right) + X \cos \theta_\ell \right]~~,
\eqno(2.4e)
$$
where $X$ and $\beta$ are defined in Appendix A, and our convention is
$\epsilon_{0123}=1$.

  In the laboratory frame (the $B$-frame), the above relations become
$$
P_B \cdot p = m_B E_p ~~,
\eqno(2.5a)
$$
$$
P_B \cdot q = m_B E_q ~~,
\eqno(2.5b)
$$
$$
P_B \cdot p_\ell = m_B E_\ell ~~.
\eqno(2.5c)
$$
The four-body phase space element is given by
$$
\begin{array}{rcl}
d (PS) & = & \frac{d^3p}{(2\pi)^3 2E_p} \frac{d^3q}{(2\pi)^3 2E_q}
\frac{d^3p_\ell}{(2\pi)^3 2E_\ell} \frac{d^3p_\nu}{(2\pi)^3 2E_\nu}  \\
& & \times (2\pi)^4 \delta^4 (P_B - p - q - p_\ell - p_\nu) ~~.
\end{array}
\eqno(2.6)
$$

\noindent To reduce the above expression to a five-dimensional integral, we
insert the factors
$$
1 = \int \frac{d^3P}{2E_P} d s_M \delta^4 (P-p-q) ~~,~~~E_P=\sqrt{\vec{P}^2+
s_M},\eqno(2.7a)
$$
$$
1 = \int \frac{d^3 L}{2E_L} d s_L \delta^4 (L -p_\ell -p_\nu) ~~,~~~E_L=\sqrt
{\vec{L}^2+s_L},\eqno(2.7b)
$$

\noindent then

$$
d(PS) = d s_M ds_L \frac{1}{(2\pi)^8} I_M I_L I_B ~~,
\eqno(2.8)
$$

\noindent where

$$
\begin{array}{rcl}
I_M & = & \int \frac{d^3p}{2E_p} \frac{d^3q}{2E_q} \delta^4 (P-p-q) \\
& = & \frac{\pi}{4} \beta d \cos \theta ~~,
\end{array}
\eqno(2.9a)
$$

$$
\begin{array}{rcl}
I_L & = & \int \frac{d^3 p_\ell}{2E_\ell} \frac{d^3p_\nu}{2E_\nu} \delta^4
(L-p_\ell - p_\nu) \\
& = & \frac{1}{8} d \cos \theta_\ell d \phi ~~,
\end{array}
\eqno(2.9b)
$$

$$
\begin{array}{rcl}
I_B & = & \int \frac{d^3P}{2E_P} \frac{d^3L}{2E_L} \delta^4 (P_B - P - L)
\\
& = & \pi \frac{X}{m^2_B} ~~.
\end{array}
\eqno(2.9c)
$$
Finally, we obtain the desired result

$$
d(PS) = \frac{1}{2(4\pi)^6 m^2_B} X \beta ds_M ds_L d\cos
\theta d\cos \theta_\ell d \phi ~~.
\eqno(2.10)
$$

The region of integration is specified by

$$
0 < s_L < (m_B - \sqrt{s_M})^2 ~~,
\eqno(2.11a)
$$
$$
(m + m_\pi)^2 < s_M < m^2_B ~~,
\eqno(2.11b)
$$
$$
0 < \theta~, \theta_\ell < \pi ~~,
\eqno(2.11c)
$$
$$
0 < \phi < 2 \pi ~~.
\eqno(2.11d)
$$

\noindent In the next two sections we will consider the single particle
energy spectra in the laboratory frame (the $B$-frame).  For this purpose
we can make use of the relations (2.4) and (2.5) to change variables from
$s_M$ and $s_L$ to $E_p$ and $E_\ell$ or $E_q$ and $E_\ell$.  For example,

$$
ds_M ds_L = \left[ \frac{\partial (E_p ~, E_\ell)}{\partial (s_M ~, s_L)}
\right]^{-1} d E_p d E_\ell ~~.
\eqno(2.12)
$$

\noindent The Jacobian in (2.12) can be computed from (2.4) and (2.5).
Other expressions for the four-body phase space element which were useful
in doing numerical calculations and verifying results are found in [18].

\nicebreak

\noindent{\bf III.~~The Semileptonic Decay $\bar{B} \to D + \pi +
\ell \bar{\nu}$}

\bigskip

The general formalism for this type of decay has been worked out by several
authors [3,10-12].  The hadronic matrix element for this process contains
four form factors.  We will not repeat the analysis here.  In an
earlier work [3] we have shown that the combined heavy quark symmetry and
chiral symmetry requires only a single form factor to describe the decay,
provided the emitted pion is soft.

The essential results are summarized below.  The effective Lagrangian for
semileptonic weak decays is given by

$$
{\cal {L}}_{\rm eff} = \frac{G_F}{\sqrt{2}} J^\mu j_\mu ~~,
\eqno(3.1)
$$

\noindent where $G_F$ is the Fermi coupling constant, $J_\mu$ the charged
hadronic weak current and $j_\mu$ the lepton's charged weak current.  To be
specific, let us consider $\bar{B}^0 \to D^+ + \pi^0 + e^-
\bar{\nu}_e$.  We have followed the convention that a $\bar{B}$ meson
contains a $b$ quark, while a $D$ meson contains a $c$ quark.  The matrix
element is given by

$$
M_{fi} = \frac{G_F}{\sqrt{2}}V_{cb} \langle\pi^0(q) D^+(p) \mid J^{cb}_\mu \mid
\bar{B}^0(P_B)\rangle \bar{u} (p_\ell) \gamma^\mu (1 - \gamma_5) v  (p_\nu)~~,
\eqno(3.2)
$$

\noindent where $V_{cb}$ is the CKM matrix element [19] for $b \to c$
transitions.  We will write

$$
\langle\pi^0 (q) D^+ (p) \mid J^{cb}_\mu \mid \bar{B}^0 (P_B) \rangle
= \frac{if}{2f_\pi} \sqrt{m_B m_D} C_{cb} \xi H_\mu ~~,
\eqno(3.3)
$$

\noindent where the pion decay constant $f_\pi = 93$ MeV, $f$ is the
$D^\ast D\pi$ coupling constant, $\xi$ is the universal Isgur-Wise function
normalized to

$$
\xi (v \cdot v^\prime) = 1 ~~~~ {\rm at} ~~~v=v^\prime ~~,
\eqno(3.4)
$$

\noindent and $C_{cb}$ is the QCD correction factor

$$
C_{cb} (v \cdot v^\prime) = \left[ \frac{\alpha_s (m_b)}{\alpha_s (m_c)}
\right]^{- \frac{6}{25}} \left[ \frac{\alpha_s (m_c)}{\alpha_s (\mu)}
\right]^{a_L (v \cdot v^\prime)} ~~,
\eqno(3.5a)
$$

$$
a_L(w) = \frac{8}{27} \left[ \frac{w}{\sqrt{w^2 -1}} \ln \left( w +
\sqrt{w^2-1} \right) -1 \right] ~~.
\eqno(3.5b)
$$

\noindent The quantity $H_\mu$ can be extracted from the result in
[3]; it is given by

$$
H_\mu = w_1 v_\mu + w_2 v^\prime_\mu + r q_\mu + i h\epsilon_{\mu \nu
\lambda \kappa} q^\nu v^{\prime \lambda} v^\kappa ~~,
\eqno(3.6)
$$

\noindent where the four-velocities $v$ and $v^\prime$ are defined by

$$
P_B = m_B v ~~, ~~~~ p = m_D v^\prime ~~.
\eqno(3.7)
$$

\noindent The form factors $w_{1,2}$, $r$ and $h$ are explicitly known in
the soft pion limit:

$$
w_1 = \frac{q \cdot (v + v^\prime)}{2v \cdot q + 2 \Delta_B} ~~,
\eqno(3.8a)
$$

$$
w_2 = -\frac{q \cdot (v + v^\prime)}{2v^\prime \cdot q - 2 \Delta_D} ~~,
\eqno(3.8b)
$$

$$
r = - \left( 1 + v \cdot v^\prime \right) \left[ \frac{1}{2v \cdot q + 2
\Delta_B} - \frac{1}{2 v^\prime \cdot q - 2 \Delta_D} \right] ~~,
\eqno(3.8c)
$$

$$
h = \frac{1}{2v \cdot q + 2 \Delta_B} - \frac{1}{2 v^\prime \cdot q - 2
\Delta_D} ~~,
\eqno(3.8d)
$$

$$
\Delta_B \equiv m_{B^\ast} - m_B ~~, ~~~~ \Delta_D \equiv m_{D^\ast} -
m_D~~.
\eqno(3.8e)
$$

In the numerical calculations, we have incorporated the finite width of
$D^*$, $\Gamma_{D^*}$, to properly handle the $D^*$ resonance by
making the replacement:

$$
\frac{1}{2 v^\prime \cdot q - 2 \Delta_D} \to \frac{m_{D^\ast}}{(p+q)^2 -
m^2_{D^\ast} + i m_{D^\ast} \Gamma_{D^\ast}} ~~.
\eqno(3.9)
$$

\noindent The differential decay rate is then

$$
\begin{array}{lcl}
d\Gamma \left( \bar{B}^0 \to D^+ + \pi^0 + e^- \bar{\nu}_\ell
\right) & = & \frac{1}{2m_B} \mid M_{fi} \mid^2 d(PS) \\
& = & \frac{G^2_F
m_D}{8 m^2_B (4\pi)^6} \left( \frac{f}{f_\pi} C_{cb} \xi \right)^2
\mid V_{cb} \mid^2
\left( \frac{1}{4} H_\mu H^\ast_\nu L^{\mu \nu} \right)  \\
& \times & \beta Xds_M ds_L d \cos \theta d \cos \theta_\ell d \phi ~~,
\end{array}
\eqno(3.10)
$$

\noindent where the lepton tensor $L_{\mu \nu}$ is given by

$$
L_{\mu \nu} = 4  (L_\mu L_\nu - N_\mu N_\nu - s_L g_{\mu \nu} - i
\epsilon_{\mu \nu \lambda \kappa} L^\lambda N^\kappa ) ~~.
\eqno(3.11)
$$
For a charged
pion in the final state, the above expression (3.10) for $d\Gamma$ has to be
multiplied by 2 due to isospin.  A straightforward but tedious calculation
gives

$$
\begin{array}{l}
\frac{1}{4} H_\mu H^\ast_\nu L^{\mu \nu} = \mid w_1 v \cdot L + w_2
v^\prime \cdot L + r q \cdot L \mid^2 \\
- \mid w_1 v \cdot N + w_2 v^\prime \cdot N + r q \cdot N - i \frac{h}{2m_B
m_D} \sigma \mid^2 \\
- s_L \left\{ \mid w_1 \mid^2 + (w_1 w^\ast_2 + w^\ast_1 w_2) v \cdot
v^\prime + (w_1 r^\ast + w^\ast_1 r) v \cdot q \right. \\
+ \mid w_2 \mid^2 + (w_2 r^\ast + w^\ast_2 r) v^\prime \cdot q + \mid r
\mid^2 m^2_\pi + \\
+ \left. \mid h \mid^2 \left[ m^2_\pi \left( (v \cdot v^\prime)^2 -1
\right) - 2 (v \cdot q)(v^\prime \cdot q)(v \cdot v^\prime) + (q \cdot v)^2
+ (q \cdot v^\prime)^2 \right] \right\} \\
-i \left\{ \frac{1}{2m_B m_D} (w_1 w^\ast_2 - w^\ast_1 w_2) -
\frac{1}{2m_B} (w_1 r^\ast - w^\ast_1 r) - \frac{1}{2m_D} (w_2 r^\ast -
w^\ast_2 r) \right\} \sigma  \\
+ (hw^\ast_1 + h^\ast w_1) \left[ - (v \cdot q) (L \cdot v^\prime) (N \cdot
v) - (v \cdot v^\prime) (L \cdot v) (N \cdot q) \right. \\
\left. - (L \cdot q)(N\cdot v^\prime) + (v \cdot v^\prime)(L \cdot q)(N
\cdot v) + (L \cdot v^\prime) (N \cdot q) + (v \cdot q) (L \cdot v) (N
\cdot v^\prime) \right] \\
+ (h w^\ast_2 + h^\ast w_2) \left[ - (v^\prime \cdot q) (L \cdot v^\prime)
(N \cdot v) - (L \cdot v) (N \cdot q) - (v \cdot v^\prime) (L \cdot q) (N
\cdot v^\prime) \right. \\
\left. + (L \cdot q) (N \cdot v) + (v \cdot v^\prime)(L \cdot v^\prime) (N
\cdot q) + (v^\prime \cdot q) (L \cdot v) (N \cdot v^\prime) \right] \\
+ (h r^\ast + h^\ast r) \left[ -m^2_\pi (L \cdot v^\prime) (N \cdot v) - (q
\cdot v^\prime) (L \cdot v) (N \cdot q) \right. \\
- (q \cdot v) (L \cdot q) (N \cdot v^\prime) + (q \cdot v^\prime) (L \cdot
q) (N \cdot v) + (q \cdot v) ( L\cdot v^\prime) (N \cdot q) \\
\left. + m^2_\pi (L \cdot v) (N \cdot v^\prime) \right] ~~.
\end{array}
\eqno(3.12)
$$where $\sigma$ is the pseudoscalar defined by (2.4b).  In deriving
(3.12) we have also made use of the relations

$$
\epsilon_{\mu \nu \lambda \kappa} N^\mu q^\nu v^{\prime \lambda} v^\kappa =
- \frac{1}{2m_B m_D} \sigma ~~,
\eqno(3.13a)
$$

$$
\epsilon_{\mu \nu \lambda \kappa} L^\mu N^\nu v^\lambda v^{\prime \kappa} =
\frac{1}{2m_B m_D} \sigma ~~,
\eqno(3.13b)
$$

$$
\epsilon_{\mu \nu \lambda \kappa} v^{\prime \mu} q^\nu L^\lambda N^\kappa =
- \frac{1}{2m_D} \sigma ~~,
\eqno(3.13c)
$$

$$
\epsilon_{\mu \nu \lambda \kappa} v^\mu q^\nu L^\lambda N^\kappa =
- \frac{1}{2m_B} \sigma ~~.
\eqno(3.13d)
$$
With the help of (2.4), (2.5) and (2.12), we can work out the
single-particle spectra $\frac{d\Gamma}{dE_D}$, $\frac{d\Gamma}{dE_\pi}$
and $\frac{d\Gamma}{dE_\ell}$.  They will be discussed in Section V.

\nicebreak

\noindent{\bf IV.~~The Semileptonic Decay $\bar{B} \to D^\ast + \pi
+ \ell \bar{\nu}$}

\bigskip

The kinematics for this decay is very similar to the one discussed in
previous sections.  The new feature here is the polarization of the vector
meson $D^\ast$ which we will exploit in our study.

In analogy with (3.3) for $\bar{B} \to D\pi\ell\bar{\nu}$ we define
$$
\langle D^{\ast +} (p) \pi^0 (q) \mid J^{cb}_\mu \mid \bar{B}^0 (P_B) \rangle
= - \frac{i}{2} \sqrt{m_B m_{D^\ast}} \frac{f}{f_\pi} C_{cb} \xi
H^\prime_\mu ~~,
\eqno(4.1)
$$
where an extra minus sign is introduced here compared with
(3.3) due to the scalar products involving the polarization vector
$\varepsilon_\mu$ of $D^\ast$.  Heavy quark symmetry and chiral
dynamics have a definite prediction for $H^\prime_\mu$ when the emitted pion
is soft.  In this limit we find from [3]

$$
\begin{array}{rcl}
H^\prime_\mu &  = & a_1 v_\mu + a_2 v^\prime_\mu + a_3 q_\mu + a_4
\varepsilon^\ast_\mu  \\
& + & i \epsilon_{\mu \nu \lambda \kappa} \varepsilon^{\ast \nu} \left(
b_1q^\lambda v^{\prime \kappa} + b_2 q^\lambda v^\kappa  + b_3 v^\lambda
v^{\prime \kappa}  \right) ~~,
\end{array}
\eqno(4.2)
$$

\noindent where

$$
P_B = m_B v ~~, ~~~~~~~~~ p= m_{D^\ast} v^\prime ~~,
\eqno(4.3)
$$

\noindent and

$$
a_1 = - \left[ \frac{1}{- 2v \cdot q - 2 \Delta_B} + \frac{1}{2 v^\prime
\cdot q + 2 \Delta_D} \right] (\varepsilon^\ast \cdot q) ~~,
\eqno(4.4a)
$$

$$
\begin{array}{rcl}
a_2 & = & \frac{1}{-2v \cdot q - 2 \Delta_B} \left[ (\varepsilon^\ast \cdot
v) (v \cdot q) - \varepsilon^\ast \cdot q \right] \\
& & - \frac{1}{2 v^\prime \cdot q} \left[ i \epsilon_{\rho \sigma
\lambda \kappa} q^\rho \varepsilon^{\ast \sigma} v^{\prime \lambda}
v^\kappa + (q \cdot v^\prime) (\varepsilon^\ast \cdot v) \right] \\
& & - \frac{1}{2v^\prime \cdot q + 2 \Delta_D} (\varepsilon^\ast \cdot q) ~~,
\end{array}
\eqno(4.4b)
$$

$$
a_3 = \left[ \frac{1}{-2v \cdot q - 2 \Delta_B} + \frac{1}{2v^\prime \cdot
q} \right] (\varepsilon^\ast \cdot v) ~~,
\eqno(4.4c)
$$

$$
\begin{array}{rcl}
a_4 & = & \frac{1}{-2v \cdot q - 2\Delta_B} \left[ (q \cdot v^\prime) - (v
\cdot v^\prime) (v \cdot q) \right] \\
& & - \frac{1}{2 v^\prime \cdot q} \left[ (q \cdot v) - (v \cdot v^\prime)
(v^\prime \cdot q) \right] ~~,
\end{array}
\eqno(4.4d)
$$

$$
b_1 = \frac{1}{-2v \cdot q -2 \Delta_B} + \frac{1}{2 v^\prime \cdot q} (1 +
v \cdot v^\prime ) ~~,
\eqno(4.4e)
$$

$$
b_2 = \frac{1}{-2v \cdot q - 2 \Delta_B} ~~,
\eqno(4.4f)
$$

$$
b_3 = \frac{-(q \cdot v)}{-2v \cdot q - 2 \Delta_B} ~~.
\eqno(4.4g)
$$
In our numerical calculations for this process we will not employ the full
propagator (3.9) as we do for the decay $\bar{B}\to D \pi \ell\bar{\nu}$,
since none of the intermediate states can become real here.

The absolute value squared of the matrix element involves

$$
\begin{array}{rcl}
\frac{1}{4} H^\prime_\mu H^{\prime \ast}_\nu L^{\mu \nu} & = & \mid
H^\prime \cdot L \mid^2 - \mid H^\prime \cdot N \mid^2 - s_L H^\prime \cdot
H^{\prime \ast} \\
& & - i \epsilon^{\mu \nu \lambda \kappa} H^\prime_\mu H^{\prime \ast}_\nu
L_\lambda N_\kappa ~~,
\end{array}
\eqno(4.5)
$$

\noindent where the lepton tensor $L_{\mu \nu}$ is the same as
the one given by
(3.11).  Each term in (4.5) is straightforward to compute, though it is
tedious sometimes.

To begin with, we introduce the quantities

$$
\sigma_1 = \epsilon_{\mu \nu \lambda \kappa} N^\mu \varepsilon^{\ast \nu}
q^\lambda v^{\prime \kappa} ~~,
\eqno(4.6a)
$$
$$
\sigma_2 = \epsilon_{\mu \nu \lambda \kappa} N^\mu \varepsilon^{\ast \nu}
q^\lambda v^\kappa ~~,
\eqno(4.6b)
$$
$$
\sigma_3 = \epsilon_{\mu \nu \lambda \kappa} N^\mu \varepsilon^{\ast \nu}
v^\lambda v^{\prime \kappa} ~~,
\eqno(4.6c)
$$
$$
\sigma_4 = \epsilon_{\mu \nu \lambda \kappa} v^\mu \varepsilon^{\ast \nu}
q^\lambda v^{\prime \kappa} ~~.
\eqno(4.6d)
$$
and
$$
\sigma^\prime_1 = \epsilon_{\mu \nu \lambda \kappa} \varepsilon^{\ast \mu}
\varepsilon^\nu q^\lambda v^{\prime \kappa} ~~,
\eqno(4.7a)
$$
$$
\sigma^\prime_2 = \epsilon_{\mu \nu \lambda \kappa} \varepsilon^{\ast \mu}
\varepsilon^\nu q^\lambda v^\kappa ~~,
\eqno(4.7b)
$$
$$
\sigma^\prime_3 = \epsilon_{\mu \nu \lambda \kappa} \varepsilon^{\ast \mu}
\varepsilon^\nu v^\lambda v^{\prime \kappa} ~~,
\eqno(4.7c)
$$
$$
\sigma^\prime_4 = \epsilon_{\mu \nu \lambda \kappa} \varepsilon^{\ast \mu}
\varepsilon^\nu L^\lambda N^\kappa ~~.
\eqno(4.7d)
$$

\noindent Then, we find

$$
\begin{array}{rcl}
H^\prime \cdot L & = & a_1 v \cdot L + a_2 v^\prime \cdot L + a_3 q \cdot L
+ a_4 \varepsilon^\ast \cdot L \\
& & + i \sigma_4 (m_B b_1 + m_{D^\ast} b_2 + b_3) ~~,
\end{array}
\eqno(4.8a)
$$

$$
\begin{array}{rcl}
H^\prime \cdot N & = & a_1 v \cdot N + a_2 v^\prime \cdot N + a_3 q \cdot
N + a_4 \varepsilon^\ast \cdot N \\
& & + i (\sigma_1 b_1 + \sigma_2 b_2 + \sigma_3 b_3 ) ~~,
\end{array}
\eqno(4.8b)
$$

$$
H^\prime \cdot H^{\prime \ast} = T_a + T_{ab} + T_b ~~,
\eqno(4.8c)
$$

$$
\begin{array}{rcl}
T_a & = & \mid a_1 \mid^2 + \mid a_2 \mid^2 + \mid a_3 \mid^2 m^2_\pi -
\mid a_4 \mid^2 + (a_1 a_2^\ast + a_1^\ast a_2) v \cdot v^\prime \\
& & +(a_1 a_3^\ast + a^\ast_1 a_3) v \cdot q + (a_2 a^\ast_3 + a^\ast_2
a_3) v^\prime \cdot q  \\
& & + a_1 a^\ast_4 \varepsilon \cdot v + a^\ast_1 a_4 \varepsilon^\ast
\cdot v + a_3 a^\ast_4 \varepsilon \cdot q + a^\ast_3 a_4 \varepsilon^\ast
\cdot q ~~,
\end{array}
\eqno(4.8d)
$$

$$
\begin{array}{rcl}
T_{ab} & = & -i a_1 b^\ast_1 \sigma^\ast_4 + i a^\ast_1 b_1 \sigma_4 + i
a_2 b^\ast_2 \sigma^\ast_4 - i a^\ast_2 b_2 \sigma_4 \\
& & + i a_3 b^\ast_3 \sigma^\ast_4 - i a^\ast_3 b_3 \sigma_4 \\
& & - i (a_4 b^\ast_1 + a^\ast_4 b_1) \sigma^\prime_1 - i (a_4 b^\ast_2  +
a^\ast_4 b_2) \sigma^\prime_2 \\
& & -i (a_4 b^\ast_3 + a^\ast_4 b_3) \sigma^\prime_3 ~~,
\end{array}
\eqno(4.8e)
$$

$$
\begin{array}{rcl}
T_b & = & \mid b_1 \mid^2 \left[ m^2_\pi + \mid \varepsilon \cdot q
\mid^2 - (q \cdot v^\prime )^2 \right] \\
& & + \mid b_2 \mid^2 \left[ m^2_\pi - (\varepsilon^\ast \cdot q)
(\varepsilon \cdot v) (q \cdot v) - (\varepsilon^\ast \cdot v) (\varepsilon
\cdot q) (q \cdot v) + \mid \varepsilon \cdot q \mid^2 \right. \\
& & \left. ~~~~~~~ + | \varepsilon \cdot v|^2
m^2_\pi - (q \cdot v)^2 \right] \\
& & + \mid b_3 \mid^2 \left[ 1 + \mid \varepsilon \cdot v \mid^2 - (v \cdot
v^\prime)^2 \right] \\
& & + b_1 b^\ast_2 \left[ m^2_\pi (v \cdot v^\prime) - (\varepsilon^\ast
\cdot v)(\varepsilon \cdot q) (v^\prime \cdot q) + \mid \varepsilon^\ast
\cdot q \mid^2 (v \cdot v^\prime) - (q \cdot v)(q \cdot v^\prime) \right]
\\
& & + b^\ast_1 b_2 \left[ m^2_\pi (v \cdot v^\prime) - (\varepsilon \cdot
v) (\varepsilon^\ast \cdot q) (v^\prime \cdot q) + \mid \varepsilon^\ast
\cdot q \mid^2 (v \cdot v^\prime) - (q \cdot v) (q \cdot v^\prime) \right]
\\
& & + b_2 b^\ast_3 \left[ (q \cdot v) (v \cdot v^\prime) - \mid \varepsilon
\cdot v \mid^2 (q \cdot v^\prime) + (\varepsilon^\ast \cdot v) (\varepsilon
\cdot q) (v \cdot v^\prime) - (q \cdot v^\prime) \right] \\
& & + b^\ast_2 b_3 \left[ (q \cdot v) (v \cdot v^\prime) - \mid \varepsilon
\cdot v \mid^2 (q \cdot v^\prime) + (\varepsilon \cdot v) (
\varepsilon^\ast \cdot q) (v \cdot v^\prime) - (q \cdot v^\prime) \right]
\\
& & + b_3 b^\ast_1 \left[ (q \cdot v) + (\varepsilon^\ast \cdot q)
(\varepsilon \cdot v) - (q \cdot v^\prime) (v \cdot v^\prime) \right] \\
& & + b^\ast_3 b_1 \left[ (q \cdot v) + (\varepsilon \cdot q)
(\varepsilon^\ast \cdot v) - (q \cdot v^\prime) (v \cdot v^\prime)
\right]~~.
\end{array}
\eqno(4.8f)
$$

Similarly, we write

$$
i \epsilon^{\mu \nu \lambda \kappa} H^\prime_\mu H^{\prime \ast}_\nu
L_\lambda N_\kappa = R_a + R_{ab} + R_b ~~,
\eqno(4.9a)
$$

\noindent where

$$
\begin{array}{rcl}
R_a & = & i \mid a_4 \mid^2 \sigma^\prime_4 + {\rm Im} \Bigl\{ \sigma \left( -
\frac{1}{m_B m_{D^\ast}} a_1 a^\ast_2 + \frac{1}{m_B} a_1 a^\ast_3 +
\frac{1}{m_{D^\ast}} a_2 a^\ast_3 \right) \\
& & - 2 \sigma^\ast_1 \left( a_2 a^\ast_4 - m_{D^\ast} a_3 a^\ast_4
\right) - 2 \sigma^\ast_2 \left( m_B a_3 a^\ast_4 + a_1 a^\ast_4 \right) \\
& & + 2 \sigma^\ast_3 \left( m_{D^\ast} a_1 a^\ast_4 + m_B a_2
a^\ast_4 \right) \Bigr\}
\end{array}
\eqno(4.9b)
$$

$$
\begin{array}{rcl}
R_{ab} & = & 2 {\rm Re} \left\{ a^\ast_1 b_1  \left[ (v \cdot
\varepsilon^\ast)(L\cdot q) (N \cdot v^\prime) + (N \cdot \varepsilon^\ast)
(v \cdot q) (L \cdot v^\prime) + (L \cdot \varepsilon^\ast) (N \cdot q) (v
\cdot v^\prime) \right. \right. \\
& & \left. - (L \cdot \varepsilon^\ast) (v \cdot q) (N \cdot v^\prime) - (N
\cdot \varepsilon^\ast) (L \cdot q) (v \cdot v^\prime) -
(v \cdot \varepsilon^\ast)(N \cdot q) (L \cdot v^\prime) \right] \\
& & +a^\ast_1 b_2  \left[ (v \cdot \varepsilon^\ast) (L \cdot q) (N \cdot
v) + (N \cdot \varepsilon^\ast) (v \cdot q) (L \cdot v) + (L \cdot
\varepsilon^\ast) (N \cdot q) \right. \\
& & \left. - (L \cdot \varepsilon^\ast) (v \cdot q) (N \cdot v) - (N
\cdot \varepsilon^\ast) (L \cdot q) - (v \cdot \varepsilon^\ast) ( N \cdot
q) (L \cdot v) \right] \\
& & +a^\ast_1 b_3  \left[ (v \cdot \varepsilon^\ast)(L \cdot v) (N \cdot
v^\prime) + (N \cdot \varepsilon^\ast) (L \cdot v^\prime) + (L \cdot
\varepsilon^\ast) (N \cdot v) (v \cdot v^\prime) \right. \\
& & \left. - (L \cdot \varepsilon^\ast) (N \cdot v^\prime) - (N \cdot
\varepsilon^\ast)(L \cdot v) (v \cdot v^\prime) - (v \cdot
\varepsilon^\ast) (N \cdot v) (L \cdot v^\prime) \right] \\
& & +a^\ast_2 b_1  \left[ (N \cdot \varepsilon^\ast) (v^\prime \cdot q) (L
\cdot v^\prime) + (L \cdot \varepsilon^\ast) (N \cdot q) \right. \\
& &  \left. - (L \cdot \varepsilon^\ast) (v^\prime \cdot q) (N \cdot
v^\prime) - (N \cdot \varepsilon^\ast) (L \cdot q) \right] \\
& & +a^\ast_2 b_2  \left[ (N \cdot \varepsilon^\ast) (v^\prime \cdot q) (L
\cdot v) + (L \cdot \varepsilon^\ast) (N \cdot q) (v \cdot v^\prime) - (L
\cdot \varepsilon^\ast) (v^\prime \cdot q) (N \cdot v) \right. \\
& & \left. - (N \cdot \varepsilon^\ast) (L \cdot q) (v \cdot v^\prime)
\right] \\
& & +a^\ast_2 b_3  \left[ (N \cdot \varepsilon^\ast) (v \cdot v^\prime) (L
\cdot v^\prime) + (L \cdot \varepsilon^\ast) (N \cdot v) - ( L \cdot
\varepsilon^\ast) (v \cdot v^\prime) (N \cdot v^\prime) \right. \\
& & \left. - (N \cdot \varepsilon^\ast) (L \cdot v) \right] \\
& & +a^\ast_3 b_1  \left[ (q \cdot \varepsilon^\ast) (L \cdot q) (N \cdot
v^\prime) + (N \cdot \varepsilon^\ast) m^2_\pi (L \cdot v^\prime) + (L \cdot
\varepsilon^\ast) (N \cdot q) (q \cdot v^\prime) \right. \\
& &  \left. - (L \cdot \varepsilon^\ast) m_\pi^2 (N \cdot v^\prime) - (N
\cdot \varepsilon^\ast) (L \cdot q) (q \cdot v^\prime) - (q \cdot
\varepsilon^\ast) (N \cdot q) (L \cdot v^\prime) \right] \\
& & +a^\ast_3 b_2  \left[ (q \cdot \varepsilon^\ast)(L \cdot q) (N \cdot
v) + (N \cdot \varepsilon^\ast) m_\pi^2 (L \cdot v) + (L \cdot
\varepsilon^\ast) (N \cdot q) (q \cdot v) \right. \\
& &  \left. - (L \cdot \varepsilon^\ast) m_\pi^2 (N \cdot v) - (N \cdot
\varepsilon^\ast) (L \cdot q) (q \cdot v) - (q \cdot \varepsilon^\ast) (N
\cdot q) (L \cdot v) \right] \\
& & +a^\ast_3 b_3  \left[ (q \cdot \varepsilon^\ast) (L \cdot v) (N \cdot
v^\prime) + (N \cdot \varepsilon^\ast) (q \cdot v) (L \cdot v^\prime) + (L
\cdot \varepsilon^\ast) (N \cdot v) (q \cdot v^\prime) \right. \\
& &  \left. - (L \cdot \varepsilon^\ast) (q \cdot v) (N \cdot v^\prime) -
(N \cdot \varepsilon^\ast) (L \cdot v) (q \cdot v^\prime) - (q \cdot
\varepsilon^\ast) (N \cdot v) (L \cdot v^\prime) \right] \\
& & +a^\ast_4 b_1  \left[ -(L \cdot q) (N \cdot v^\prime) + (N \cdot
\varepsilon^\ast) (\varepsilon \cdot q) (L \cdot v^\prime) - (L \cdot
\varepsilon^\ast) (\varepsilon \cdot q) (N \cdot v^\prime) \right. \\
& & \left. + (N \cdot q) (L \cdot v^\prime) \right] \\
& & +a^\ast_4 b_2  \left[ -(L \cdot q) (N \cdot v) + (N \cdot
\varepsilon^\ast) (\varepsilon \cdot q) (L \cdot v) + (L \cdot
\varepsilon^\ast) (N \cdot q) (\varepsilon \cdot v) \right. \\
& &  \left. - (L \cdot \varepsilon^\ast) (\varepsilon \cdot q)(N \cdot v) -
(N \cdot \varepsilon^\ast) (L \cdot q) (\varepsilon \cdot v)
+ (N \cdot q) (L \cdot v) \right] \\
& &  + a^\ast_4 b_3  \left. \left[ - (L \cdot v) (N \cdot v^\prime) + (N
\cdot \varepsilon^\ast) (\varepsilon \cdot v) (L \cdot v^\prime) - (L \cdot
\varepsilon^\ast ) (\varepsilon \cdot v) (N \cdot v^\prime) \right. \right.
\\
& & \left. \left. + (N \cdot v) (L \cdot v^\prime) \right] \right\} ~~,
\end{array}
\eqno(4.9c)
$$
and
$$
\begin{array}{rcl}
R_b & = & -{\rm Im}\Big\{ \left[ (L \cdot v^\prime) (N \cdot q) - (L \cdot q)
(N \cdot v^\prime) \right] \left( \mid b_1 \mid^2 \sigma^\prime_1 +
b^\ast_2 b_1 \sigma^\prime_2 + b^\ast_3 b_1 \sigma^\prime_3 \right) \\
& & +\left[ (L \cdot v) (N \cdot q) - (L \cdot q) (N \cdot v) \right] \left(
 \mid b_2 \mid^2 \sigma^\prime_2 +b^\ast_1 b_2 \sigma'_1 + b^\ast_3 b_2
\sigma'_3 \right) \\
& & + \left[ (L \cdot v^\prime) (N \cdot v) - (L \cdot v) (N \cdot
v^\prime) \right] \left( \mid b_3 \mid^2\sigma^\prime_3+
b^\ast_2 b_3 \sigma'_2 + b^\ast_1 b_3 \sigma^\prime_1 \right) \\
& & + \left[ (L \cdot q) (N \cdot \varepsilon^\ast) - (L \cdot
\varepsilon^\ast) (N \cdot q) \right] \left( b^\ast_1 b_2
 - b^\ast_2 b_1 \right)\sigma_4^\ast \\
& & + \left[ (L \cdot \varepsilon^\ast)(N \cdot v) - (L \cdot v) (N \cdot
\varepsilon^\ast) \right] \left( b^\ast_2 b_3  -
b^\ast_3 b_2  \right)\sigma_4^\ast \\
& & + \left[ (L \cdot v^\prime) (N \cdot \varepsilon^\ast) - (L \cdot
\varepsilon^\ast) (N \cdot v^\prime) \right] \left( b^\ast_3 b_1
 - b^\ast_1 b_3  \right)\sigma_4^\ast \Big\} ~~.
\end{array}
\eqno(4.9d)
$$

The polarization state of a massive vector meson is not a Lorentz invariant
concept.  A state with a definite polarization in one frame of reference
will become a linear combination of states with different polarizations in
another frame of reference.  When the polarization of a vector meson is
specified, we have to give the frame of reference in which it is
defined.  In our numerical calculations and results to be presented in the
next section, we will employ states of $D^\ast$ with definite polarizations
in the rest frame of the $\bar{B}$ meson (the $B$-frame).
The polarization vectors in the $M$-frame and their Lorentz transforms
in the $L$-frame are provided in Appendix A.  Polarization vectors in the
$B$-frame are not explicitly given, but they are not difficult to
construct.

\nicebreak

\noindent{\bf V.~~Results and Discussion}

\bigskip

In this section we make use of the results obtained in the last two
sections to compute the single particle spectra for the charmed meson ($D$
or $D^\ast$), the pion and the electron, and the total rates for $\bar{B}
\to D + \pi +\ell\bar{\nu}$ and $\bar{B} \to D^\ast + \pi +
\ell\bar{\nu}$.  These results are presented in a series of figures.  The
energies in the single particle spectra of these figures are those measured
in the rest frame of the $\bar{B}$ meson; the polarizations of $D^\ast$ are
also specified in the $B$-frame.
Since the validity of chiral symmetry
demands the emitted pions be soft, we must impose cutoffs on the pion
momenta in our calculation.  It is not clear how soft a pion must be for
chiral symmetry to work, nor is it obvious in which reference frame
the pion has to be soft.  For the problem at hand, there are two obvious
frames: the rest frame of the $\bar{B}$ meson and the rest frame of $D(D^\ast)$
system.  As the pion is soft, the center of mass frame of the $D(D^\ast)\pi$
system is approximately the same as the rest frame of the $D(D^\ast)$
meson.  We refer to the latter as the $D$-frame or the $D^\ast$-frame
as the case may be.  Results are presented with the pion momentum cutoff in the
$B$-frame, the $D$-frame, or both.  We simply cut off the pion's 3-momentum
at 100 MeV/c or 200 MeV/c in the appropriate frame of reference.
Comparison among the plots for different cutoffs should give some idea of
the sensitivity of our results to the different cutoff procedures.
Generally speaking, the shapes of various spectra do not differ very much.
The rate for $\bar{B} \to D\pi\ell\bar{\nu}$ does not change by much, but
that for $\bar{B} \to D^*\pi\ell\bar{\nu}$ varies by almost an order of
magnitude.

Throughout our calculations, we use the following values of the
well-measured parameters [7,20]:

$$
\begin{array}{cc}
m_{B^\pm}=5278.6~{\rm MeV}, & m_{B^0} = 5278.7 ~{\rm MeV},~~~ m_{B^\ast} =
5331.3 ~{\rm MeV}, \\
m_{D^0}=1864.5~{\rm MeV}, & m_{D^\pm} = 1869.3 ~{\rm MeV},~~~m_{D^{*\pm}} =
2010.1 ~{\rm MeV},  \\  m_{D^{*0}}=2007.1~{\rm MeV}, &
f_\pi = 93.0 ~{\rm MeV},~~~G_F = 1.16637 \times 10^{-11} ~{\rm MeV}^{-2}.
\end{array}
\eqno(5.1)
$$

\noindent We also use

$$
f = -1.5~,\quad\quad\Gamma_{D^{*+}} = 141 ~{\rm keV}~
,\quad\quad \Gamma_{D^{*0}} = 102 ~{\rm keV}~.
\eqno(5.2)
$$

\noindent In (5.2), the value for the fundamental coupling constant $f$ is
the one from the quark model given in our earlier work [3], and
the ${D^*}$ widths are our prediction [14].
We also used other values for $\Gamma_{D^\ast}$ as shown in Figs. 7-10.
As for the Isgur-Wise form factor, we use the one given by Burdman [21]

$$
\xi (y) = 1 - \rho^2 (y-1) + c (y-1)^2 ~.
\eqno(5.3)
$$
\noindent with
$$
\rho = 1.08\pm 0.10~, \quad\quad c = 0.62\pm 0.15~.
\eqno(5.4)
$$
In his fit Burdman found
$$
|V_{cb}| = 0.041\pm 0.005\pm 0.002~,
\eqno(5.5)
$$

\noindent which is somewhat smaller than the values obtained by several
experimental analyses [22].  But it is not easy to compare (5.5) with other
analyses since Burdman's fit includes QCD corrections.  As a result,
$\xi(y)$ given by (5.3) is only applicable to $\bar{B} \to D (D^\ast)$
decays.

We now consider some details of the decays $\bar{B}^0 \to D+
\pi+\ell\bar{\nu}$.  Since the intermediate $D^\ast$ can be on its mass shell,
results for this decay rate are separated into two categories: resonant and
nonresonant.  The resonant part is defined as those events with the
invariant mass of $D\pi$ satisfying [23]:

$$
\mid m (D\pi) - m_{D^\ast} \mid < 3 \Gamma_{D^\ast} ~.
\eqno(5.6)
$$

\noindent All others are nonresonant.  For comparison with experiment,
the resonant part
is identified as $\bar{B}^0 \to D^*\ell\bar{\nu}$ followed by the decay
of the $D^*$ to the specific $D\pi$ state, while the nonresonant part is
identified as $\bar{B}^0 \to D\pi\ell\bar{\nu}$.  We have found that
the contribution
from the $D^\ast$ pole dominates both the resonant and nonresonant decays.
Since the most important Feynman diagram is Fig. 1b, where the
pion is emitted in a transition from $D^*$ to $D$, it is most reasonable
to cut the pion's three momentum off in the $D$-frame.
We see from Figs. 4-6 that the shapes do not change much
between the single-particle spectra in the resonant region and their
counterparts in the nonresonant region.  However, the rates in the resonant
region are larger than those in the nonresonant region by a factor of 7.
When the cutoff increases
from 100 MeV to 200 MeV, the nonresonant rate increases by about 15\%.
The pion momentum cutoff of 100 MeV or 200 MeV in the $D$-frame has no effect
on the resonant contribution, since the pion momentum in this frame
is only about 40 MeV.  We have
investigated the sensitivity to $\Gamma_{D^\ast}$ of the shapes of the
single particle spectra.  They hardly change as $\Gamma_{D^\ast}$ varies
from 0.1 MeV to 1 MeV.

To estimate the branching ratios of $\bar{B}^0\to(D\pi)_{\rm res}+e^-\bar{\nu}
_e$ and $\bar{B}^0\to (D\pi)_{\rm nonres}+e^-\bar{\nu}_e$,
we first convert the mean lifetime of $\bar{B}$ mesons [7]

$$
\tau_B = (12.9 \pm 0.5) \times 10^{-13} s~,
\eqno(5.7)
$$

\noindent to a total decay width of

$$
\Gamma_B  = \frac{\hbar}{\tau} = (0.51 \pm 0.02 ) \times 10^{-9} ~{\rm MeV}~.
\eqno(5.8)
$$
The width $\Gamma_{D^*}$ also affects the total decay rates for $\bar{B}^0
\to (D\pi)_{\rm res}+e^-\bar{\nu}_e$ and $\bar{B}^0\to (D\pi)_{\rm nonres}
+e^-\bar{\nu}_e$. The dependence of the integrated rates on $\Gamma_{D^*}$
is displayed in Figs. 7-10. We have fixed the value of
$f$ in the amplitudes by (5.2), but have treated $\Gamma_{D^*}$ as a free
parameter in the $D^\ast$ propagator (3.9). In this way, a linear
relationship between the integrated rates and $1/\Gamma_{D^*}$ is expected
theoretically. The details are presented in Appendix B.
We notice that as $\Gamma_{D^*}$ varies from 0.1 MeV to 1 MeV,
$\Gamma[\bar{B}^0\to (D^+\pi^0)_{\rm res}+e^-\bar{\nu}_e]$ decreases from
about $1.2\times 10^{-11}$ MeV to $1.2\times 10^{-12}$ MeV for
$q_{\rm max}=100$
MeV or $q_{\rm max}=200$ MeV. For $q_{\rm max}=100$ MeV, the corresponding
change for the rate $\Gamma[\bar{B}^0\to (D^+\pi^0)_{\rm nonres}+e^-\bar{\nu}
_e]$ is from $1.5\times 10^{-12}$ MeV to $2.2\times 10^{-13}$ MeV.
For $q_{\rm max}=200$ MeV, the corresponding change in the nonresonant rate
is from $1.6\times 10^{-12}$ MeV to $3.7\times 10^{-13}$ MeV.
Similar variations are also found in the
rates for $B^- \to D^0\pi^0e^-\bar{\nu}_e$ with respect to changes of
$\Gamma_{D^\ast}$ and the pion momentum cutoffs.  Because
$m_{D^{\ast0}} < m_{D^+} + m_{\pi^-}$, the rates for
$B^- \to (D^+\pi^-)_{\rm res}e^-\bar{\nu}_e$ are completely negligible.
The rates for
$B^- \to (D^+\pi^-)_{\rm nonres}e^-\bar{\nu}_e$ are nearly independent of
$\Gamma_{D^\ast}$.  See appendix B for
more discussion on $B^- \to D^+\pi^-e^-\bar{\nu}_e$.

   From Figs. 7-10, we find the decay rates with a neutral pion
$$\Gamma[\bar{B}^0\to (D^+\pi^0)_{\rm res}+e^-\bar{\nu}_e]=\,8.50\times
10^{-12}\,{\rm MeV},\eqno(5.9a)$$
$$\Gamma[\bar{B}^0\to (D^+\pi^0)_{\rm nonres}+e^-\bar{\nu}_e]=\,1.09\times
10^{-12}\,{\rm MeV},\eqno(5.9b)$$
$$\Gamma[B^-\to (D^0\pi^0)_{\rm res}+e^-\bar{\nu}_e]=\,1.78\times
10^{-11}\,{\rm MeV},\eqno(5.9c)$$
$$\Gamma[B^-\to (D^0\pi^0)_{\rm nonres}+e^-\bar{\nu}_e]=\,2.16\times
10^{-12}\,{\rm MeV}.\eqno(5.9d)$$
The corresponding branching ratios are
$$B[\bar{B}^0\to (D^+\pi^0)_{\rm res}+e^-\bar{\nu}_e]=\,1.67\%,
\eqno(5.10a)$$
$$B[\bar{B}^0\to (D^+\pi^0)_{\rm nonres}+e^-\bar{\nu}_e]=\,0.21\%,
\eqno(5.10b)$$
$$B[B^-\to (D^0\pi^0)_{\rm res}+e^-\bar{\nu}_e]=\,3.49\%,
\eqno(5.10c)$$
$$B[B^-\to (D^0\pi^0)_{\rm nonres}+e^-\bar{\nu}_e]=\,0.42\%.
\eqno(5.10d)$$
If we identify the $D\pi$'s in the resonant region with the $D^*$, Eq. (5.10a)
is the combined branching ratio for $\bar{B}^0\to D^{*+}e^-\bar{\nu}_e$
and $D^{*+}\to D^+\pi^0$.
The decay $D^{*+}\to D^+\pi^0$ is predicted to have a branching ratio of $31.2
\%$ [14], thus
$$
\begin{array}{ccl}
$$B[\bar{B}^0\to D^{\ast+}+e^-\bar{\nu}_e] & = & {B[\bar{B}^0\to
(D^+\pi^0)_{\rm res}+e^-\bar{\nu}_e]\over B[D^{*+}\to D^+\pi^0]} \\
& = & 5.35\%.
\end{array}\eqno(5.11)$$
The agreement
between (5.11) and the data (1.2) is very good. Encouraged by this success, we
would like to relate (5.10c) to the branching ratio of $B^-\to D^{*0}e^-
\bar{\nu}_e$. To do this we notice that $D^{*0}\to D^+\pi^-$ is kinematically
forbidden, and $D^{*0}$ has a substantial radiative decay. Using the
branching ratio of $66.7\%$ for $D^{*0}\to D^0\pi^0$ [14], we find
$$B[B^-\to D^{*0}+e^-\bar{\nu}_e]=\,5.23\%,\eqno(5.12)$$
which agrees with the data (1.4).  The success of the predictions (5.11)
and (5.12) represents a triumph for heavy quark symmetry; it is independent
of the chiral symmetry of light quarks.
For the nonresonant $D\pi$ final states,
we can read off from Figs. 8 and 10 the contribution from processes with a
charged pion.  Combining this with (5.9) and (5.10), we find
$$B[\bar{B}^0\to (D\pi)^+_{\rm nonres}+e^-\bar{\nu}_e]=\,0.68\%,
\eqno(5.13a)$$
$$B[B^-\to (D\pi)^0_{\rm nonres}+e^-\bar{\nu}_e]=\,0.45\%.
\eqno(5.13b)$$
We notice in passing that isospin symmetry is reasonably good for
$\bar{B}^0$ decays but not so for $B^-$ decays, as a consequence
of the fact that $m_{D^{\ast0}} < m_{D^+} + m_{\pi^-}$.
So far, the results quoted above are for $q_{\rm max}=100$ MeV in the $D^*$
frame. For $q_{\rm max}=200$ MeV, we obtain
$$B[\bar{B}^0\to (D\pi)^+_{\rm nonres}+e^-\bar{\nu}_e]=\,0.77\%,
\eqno(5.14a)$$
$$B[B^-\to (D\pi)^0_{\rm nonres}+e^-\bar{\nu}_e]=\,0.53\%.
\eqno(5.14b)$$

As mentioned in the Introduction, there appears to be a deficit between the
branching ratio for the inclusive semileptonic decays $B \to e^\pm
\nu_e$+hadrons and the sum of the two exclusive channels $B^0 \to
D^- \ell^+ \nu$ and $B^0 \to D^{\ast -} \ell^+ \nu$.  Our study
shows that the nonresonant decay $\bar{B}^0 \to D \pi\ell\bar{\nu}$
can have a substantial branching ratio, although not enough to account for
the difference.  Nevertheless,
it is measurable and is interesting in its own right.

Before we leave the subject of $\bar{B} \to (D\pi)_{\rm nonres}
\ell \bar{\nu}$, we would like to repeat a comment made in Appendix B.  The
results for nonresonant contributions (5.9b), (5.9d), (5.13) and (5.14) are
very sensitive to the definition (5.6) for resonant contributions.  To
compare our predictions with future experiments, we must bear this point in
mind.  Of course, the sum of the resonant and nonresonant contributions
is independent of this arbitrary division into resonant and
nonresonant parts.

We now turn to the decay $\bar{B}^0 \to D^{\ast +} \pi^0 e^-
\bar{\nu}_e$.   The results are shown in Figs. 11-16.  The overall rates are
smaller than $\bar{B}^0 \to D \pi \ell\bar{\nu}
$ by two or three orders of magnitude.  For this process, we have imposed
the pion momentum cutoff in the $B$-frame or in both the $B$-frame and the
$D^\ast$-frame.  We see that different cutoffs give significantly
different rates. In addition to the figures, integrated rates for
$\bar{B}^0 \to D^{\ast+}\pi^0e^-\bar{\nu}_e$ are given in Table I.

\begin{table*}[t]
\begin{center}
\begin{tabular}{lccccccc}
\multicolumn{8}{l}{Table~I. Integrated rates of $\bar{B}^0\to D^{\ast+}\pi^0
e^-\bar{\nu}_e$ with different pion momentum cutoff.} \\
\multicolumn{8}{l} {\hbox to 48 pt {\hfil} We also list
the branching ratios and the percentage contributions from} \\
\multicolumn{8}{l} {\hbox to 48 pt {\hfil} each polarization of
$D^\ast$ in the $B$-frame. The longitudinal, left-handed} \\
\multicolumn{8}{l} {\hbox to 48 pt {\hfil} and right-handed polarizations are
labelled $L_0,~L$ and $R$, respectively.} \\
\hline\hline
\noalign{\vspace{2pt}}
cutoff & pion momentum & rate (MeV) & branching & $L_0(\%)$ & $L(\%)$ &$R(\%)$
  \\
frame & cutoff (MeV) & & ratio  & & & \\
\noalign{\vspace{2pt}}
\hline
\noalign{\vspace{2pt}}
$B$ & $100$ & $3.20\times 10^{-15}$ & $0.63\times 10^{-5}$ &  26 & 57 & 17 \\
\noalign{\vspace{3pt}}
$B$ & $200$ & $1.84\times 10^{-14}$ & $0.36\times 10^{-4}$ &  30 & 52 & 18 \\
\noalign{\vspace{3pt}}
$B$ and $D^\ast$ & $100$ & $9.62\times 10^{-16}$ & $0.19\times 10^{-5}$ &  27
 & 54 & 19 \\
\noalign{\vspace{3pt}}
$B$ and $D^\ast$ & $200$ & $1.04\times 10^{-14}$ & $0.20\times 10^{-4}$ &  33
 & 48 & 19 \\
\noalign{\vspace{2pt}}
\hline\hline
\end{tabular}
\end{center}
\end{table*}

The polarization of the vector meson $D^\ast$ is a new feature of this
decay.  It can be exploited to study the nature of weak interaction
dynamics.  In Figs. 11-16, we show the single particle spectra for
each polarization of $D^\ast$ in the $B$-frame.  In all cases,
contributions from the left-handed and longitudinal
polarizations dominate that from the right-handed polarization.  This
can be simply understood as a result of the $V-A$ coupling of the quarks to
the $W^\pm$ bosons.  The charmed quark produced by the $\bar{B}$ decay is
predominantly left-handed.  The helicity of this charmed quark will not be
affected by the creation of the soft pion by the light quark
interactions.  A simple reflection will show that a left-handed charmed
quark can only lead to a vector meson $D^\ast$ with a left-handed
polarization or a longitudinal polarization.  Since the charmed quark has a
finite mass, there is some contamination from the right-handed component.
This contamination should be small when the charmed quark (hence $D^\ast$)
is energetic.  This reasoning is indeed borne out by our calculations.

Our results so far rely on a specific fit to the Isgur-Wise form factor and
a choice of the values of $f$, $V_{cb}$, etc.  All these uncertainties
will disappear if we take the ratio of the corresponding quantities at the
same value of $v \cdot v^\prime$ in the decays $\bar{B} \to D \pi
\ell\bar{\nu}$ and $\bar{B} \to D^\ast\pi\ell\bar{\nu}$.  These ratios
are the model independent predictions from the heavy quark symmetry and chiral
symmetry.  They are displayed in Figs. 17-20.  We await the day when we
will be able to compare these curves with experimental data.

To get a better idea of how the decay rates depend on the pion momentum
cutoffs, we have extrapolated our results beyond the soft pion limit.  In
Figs. 21-23, we show the various decay rates as a function of the pion
momentum cutoff in the $B$-frame, the $D$-frame, or both.  Since
$\bar{B} \to (D\pi)_{\rm nonres}\ell\bar{\nu}$ is dominated by the $D^*$
pole, it is not very sensitive to a change in cutoff.
Unfortunately, for $\bar{B} \to D^*\pi\ell\bar{\nu}$ the rates vary rapidly
with the cutoff at low pion momenta.
This exercise raises an important question:  What constitutes a soft pion?

Finally, so far we have completely neglected the contribution from $\bar{B}
\to D^{\ast \ast} \ell\bar{\nu}$, $D^{\ast \ast} \to D^\ast
\pi$.  Theoretically, this is justified since $D^{\ast \ast}$ and $D^\ast$
are nondegenerate, so the amplitude vanishes in the soft pion limit.
However, this contribution can be important in practice.  It certainly
deserves further study [24].

\bigskip\bigskip
\bigskip
\goodbreak

\centerline{\bf Acknowledgments}

\bigskip

One of us (T.M.Y.) would like to express his deep appreciation of the
hospitality extended to him by the Theory Group at the Institute of
Physics, Academia Sinica, Taipei, Taiwan, ROC during his stay there where
part of the work was done.  H.Y.C. wishes to thank Prof. C. N. Yang and the
Institute for Theoretical Physics at Stony Brook for their hospitality
during his stay there for sabbatical leave.  T.M.Y.'s work is supported in
part by the National Science Foundation.  This research is suppored in part
by the National Science Council of ROC under Contract Nos.
NSC82-0208-M001-001Y, NSC82-0208-M001-016, NSC82-0208-M001-060 and
NSC82-0208-M008-012.

\nicebreak

\noindent \centerline{{\bf Appendix A }}

\bigskip

In this appendix we present some of the properties of various four-momenta
in the three coordinate systems mentioned in the text (the $B$-, $M$-, and
$L$-frames).  Many of the results can be found in the literature [10,13,17].

In the center of mass system of $D(D^\ast)\pi$ (the $M$-frame), we have

$$
\begin{array}{lcl}
p & = & \mid \vec{p} \mid ~=~ \mid \vec{q} \mid \\
& = & \frac{1}{2} \sqrt{s_M} \beta ~,
\end{array}
\eqno(A.1)
$$

$$
L  ~=~  \mid \vec{L} \mid ~=~ \frac{X}{\sqrt{s_M}} ~~,
\eqno(A.2)
$$

\noindent where
$$X=\,\sqrt{(P\cdot L)^2-s_Ms_L},\eqno(A.3)$$
and
$$
\beta = \frac{1}{s_M} \left[ s^2_M - 2s_M \left( m^2 + m^2_\pi
\right) + \left( m^2 - m^2_\pi \right)^2 \right]^{\frac{1}{2}} ~~.
\eqno(A.4)
$$
In the $M$-frame, the components of the four-vectors $P^\mu$ and $L^\mu$ as
well as the linear polarization vectors $e^\mu_{1,2,3}$ are given in
Table II.
The values of $\mid \vec{p} \mid$ and $P \cdot L$ are given by (A.1) and
(2.3a), respectively.  The three linear polarization vectors of $D^\ast$ in
the $M$-frame as indicated in Fig. 3 are denoted by $e^\mu_1$, $e^\mu_2$,
and $e^\mu_3$.  Their components in the $M$-frame are listed in Table II.

\begin{table*}[t]
\begin{center}
\begin{tabular}{lccccccc}
\multicolumn{8}{l}{Table~II. Components of various four-vectors in the} \\
\multicolumn{8}{l} {\hbox to 48 pt {\hfil} center of mass frame of
$D(D^\ast)\pi$} \\
\multicolumn{8}{l} {\hbox to 48 pt {\hfil} (the $M$-Frame)} \\
\hline\hline
\noalign{\vspace{2pt}}
$a^\mu$ & $a^x$ & $a^y$ & $a^z$ & $a^0$ \\
\noalign{\vspace{2pt}}
\hline
\noalign{\vspace{2pt}}
$p^\mu$ & $p\sin\theta$ & $0$ & $ p\cos\theta $ & $\frac{1}{2 \sqrt{s_M}}
(s_M - m^2_\pi + m^2)$ \\
\noalign{\vspace{3pt}}
$L^\mu$ & $0$ & $0$ & $-L$ & $\frac{1}{\sqrt{s_M}} P \cdot L $ \\
\noalign{\vspace{3pt}}
$e^\mu_1$ & $\cos \theta$ & $0$ & $-\sin \theta$ & $0$ \\
\noalign{\vspace{3pt}}
$e^\mu_2$ & $0$ & $1$ & $0$ & $0$ \\
\noalign{\vspace{3pt}}
$e^\mu_3$ & $ \frac{p^0}{m} \sin \theta $ & $0$ & $ \frac{p^0}{m} \cos
\theta $ & $ \frac{p}{m} $ \\
\noalign{\vspace{2pt}}
\hline\hline
\end{tabular}
\end{center}
\end{table*}

Sometimes, we need to know the Lorentz transformations that connect the
different frames of references.  A simple calculation gives the parameters

$$
\beta_{ML} = \frac{X}{P \cdot L} ~~,
\eqno(A.5a)
$$

$$
\beta_{MB} = \frac{X}{P \cdot L + s_M} ~~.
\eqno(A.5b)
$$
If $a^\mu_L$ and $a^\mu_M$ are components of the same four-vector in the
$L$-frame and the $M$-frame, respectively, they are related by the Lorentz
transformations

$$
a^z_L = \gamma_{ML} \left( a^z_M + \beta_{ML} a^0_M \right) ~~,
\eqno(A.6a)
$$

$$
a^0_L = \gamma_{ML} \left( a^0_M + \beta_{ML} a^z_M \right)  ~~.
\eqno(A.6b)
$$

\noindent Similar equations hold using $\beta_{MB}$ to relate $a^\mu_B$
and $a^\mu_M$, the components of a vector in the $B$ and $M$ frames.

 From $Q^\mu$ and $P^\mu$ we can construct another four-momentum which is
orthogonal to $P^\mu$:

$$
Q^\prime_\mu = Q_\mu - \frac{m^2 - m^2_\pi}{s_M} P_\mu
\eqno(A.7)
$$

\noindent with

$$
Q^\prime \cdot P = 0 ~~.
\eqno(A.8)
$$
Now, we are ready to list the components of various four-vectors in
different frames of reference.  In the $L$-frame, the components of
various four-vectors are listed in Table III.

\begin{table*}[t]
\begin{center}
\begin{tabular}{lccccccc}
\multicolumn{8}{l}{Table~III. Components of various four-vectors in the rest
frame of the lepton} \\
\multicolumn{8}{l} {\hbox to 48 pt {\hfil} pair (the $L$-Frame)} \\
\hline\hline
\noalign{\vspace{2pt}}
$a^\mu$ & $a^x$ & $a^y$ & $a^z$ & $a^0$ \\
\noalign{\vspace{2pt}}
\hline
\noalign{\vspace{2pt}}
$L^\mu$ & $0$ & $0$ & $0$ & $ \sqrt{s_L}$ \\
\noalign{\vspace{3pt}}
$P^\mu$ & $0$ & $0$ & $X/\sqrt{s_L}$ & $P \cdot L/{\sqrt{s_L}}$ \\
\noalign{\vspace{3pt}}
$Q^{\prime\mu}$ & $\sqrt{s_M} \beta \sin \theta $ & $0$ & $ P \cdot L \beta
\cos \theta/\sqrt{s_L}$ & $X \beta \cos \theta/\sqrt{s_L}$ \\
\noalign{\vspace{3pt}}
$p^\mu_\ell$ & $\frac{1}{2} \sqrt{s_L} \sin \theta_\ell \cos \phi$ & $-
\frac{1}{2} \sqrt{s_L}\sin \theta_\ell \sin \phi$ & $- \frac{1}{2}
\sqrt{s_L} \cos \theta_\ell$ & $\frac{1}{2} \sqrt{s_L}$ \\
\noalign{\vspace{2pt}}
\hline\hline
\end{tabular}
\end{center}
\end{table*}

The components of $P_B$ and $p_\nu$ can be inferred from the ones given
above in the Tables.  Furthermore, with the help of a Lorentz
transformation we can determine the components of all the four-vectors in
another coordinate system.

%\vskip 1.5 cm
\nicebreak

\centerline{{\bf Appendix B}}
\bigskip

   In this appendix we derive a linear relation between the
semileptonic decay rates with a soft pion emission and the inverse of the
$D^*$ width.  In what follows, the $D^*D\pi$ coupling constant $f$
which appears in the pion emission vertices is held fixed,
while the width $\Gamma_{D^*}$ is considered to be a free parameter in the
$D^*$ propagator.

   We should emphasize that our numerical work does not make the approximation
presented below.  The purpose of this appendix is to understand the
regularities exhibited in the numerical results of Figs. 7-10.

   Consider the Feynman diagram with the $D^*$ pole contributing to $\bar{B}
\to D\pi\ell\bar{\nu}$ (Fig. 1b). The matrix element can be written as

$$M(\bar{B}\to D\pi\ell\bar{\nu})=\sum_\lambda M[D^*(\lambda)\to D\pi]{1\over
P^2-m^2+im\Gamma_{D^*}}\,M[\bar{B}\to D^*(\lambda)\ell\bar{\nu}],\eqno(B.1)$$

\noindent
where $\lambda$ denotes the polarization of $D^*$, and $P$ and $m$ are the
4-momentum and mass of $D^*$, respectively. For example,
$$M[D^*(\lambda)\to D\pi^a]=\,u^*(D^*){1\over 2}\tau^au(D)\sqrt{m_Dm}\,{f\over
f_\pi}\,\varepsilon(\lambda)\cdot q,\eqno(B.2)$$
where $q$ is the pion momentum. The decay rate for $\bar{B}\to D\pi\ell\bar{
\nu}$ due to the $D^*$ pole is

$$
\begin{array}{ccl}
\Gamma_{\rm D^*\,pole}(\bar{B}\to D\pi\ell\bar{\nu}) &=& {1\over 2m_B}\,{1\over
2\pi}\sum_\lambda\int ds_M|M[\bar{B}\to D^*(\lambda)\ell\bar{\nu}]|^2  \\
& \times & (2\pi)^4\delta^4(P_B-p_\ell-p_\nu-P){d^3p_\ell\over (2\pi)^32E_\ell}
\,{d^3p_\nu\over (2\pi)^32E_\nu}\,{d^3P\over (2\pi)^32E_P}  \\
& \times & {1\over (s_M-m^2)^2+m^2\Gamma_{D^*}^2}\int |M[D^*(\lambda)\to D\pi]
|^2  \\  &\times& (2\pi)^4\delta^4(P-q-p){d^3p\over (2\pi)^32E_p}\,{d^3q\over
(2\pi)^32E_q}.
\end{array}\eqno(B.3)$$

\noindent This equation follows from inserting the factor
$$1=\int ds_M\,{d^3P\over 2E_P}\delta^4(P-q-p),~~~E_P\equiv\sqrt{\vec{P}^2+
s_M},\eqno(B.4)$$
and the observation that one of the double sums over $D^*$ polarizations after
squaring (B.1) is eliminated when we carry out the integration over the
directions of $\vec{q}$. Making use of the standard formula for a decay
width, we find

$$\Gamma_{\rm D^*\,pole}(\bar{B}\to D\pi\ell\bar{\nu})=\,{1\over \pi}\int ds_M
{\sqrt{s_M}\,\Gamma(\bar{B}\to D^*\ell\bar{\nu},s_M)\Gamma(D^*\to D\pi,s_M)
\over (s_M-m^2)^2+m^2\Gamma_{D^*}^2}.\eqno(B.5)$$

\noindent
To obtain (B.5), we have used the fact that the width of the decay
$D^*\to D\pi$ is independent of the $D^*$ polarization.  The argument $s_M$
appears in the numerator of the integrand of (B.5) because the widths are
those appropriate for $D^*$ with a mass $\sqrt{s_M}$.

   Let us introduce the new variables
$$
s_M = m^2+ x m\Gamma_{D^*},\eqno(B.6a)$$
$$F(s_M)= \sqrt{s_M}\,\Gamma(\bar{B}\to D^*\ell\bar{\nu},s_M)\Gamma(D^*\to
D\pi,s_M).\eqno(B.6b)$$
Eq.(B.5) now becomes

$$\Gamma_{\rm D^*\,pole}(\bar{B}\to D\pi\ell\bar{\nu})=\,{1\over m\Gamma_{D^*}}
\,{1\over \pi}\int dx{1\over x^2+1}\,F[m^2+ x m\Gamma_{D^*}].\eqno(B.7)$$

\noindent
When the width $\Gamma_{D^*}$ is small, the integrand can be expanded in a
Taylor series. The leading term is

$$\Gamma_{\rm D^*\,pole}(\bar{B}\to D\pi\ell\bar{\nu})=\,{F(m^2)\over
m\Gamma_{D^*}}\,{1\over \pi}\int dx{1\over x^2+1}+\cdots.\eqno(B.8)$$

\noindent
If the range of $s_M$ is not restricted, after the $x$-integration we obtain
$$\Gamma_{\rm D^*\,pole}(\bar{B}\to D\pi\ell\bar{\nu})=\,\Gamma(\bar{B}\to D^*
\ell\bar{\nu})\,{\Gamma(D^*\to D\pi)\over\Gamma_{D^*}}+\cdots,\eqno(B.9)$$
where (B.6b) has been used. This is a well-known result in the theory of
resonances. In practice, experimental cuts are imposed on the range of $s_M$.
Suppose the cut is
$$|\sqrt{s_M}-m|<N\Gamma_{D^*},\eqno(B.10)$$
which corresponds to
$$-2N+N^2{\Gamma_{D^*}\over m}<x<2N+N^2{\Gamma_{D^*}\over m},\eqno(B.11)$$
then the region (B.10) or (B.11) is the resonant contribution and outside this
region is the nonresonant contribution. Finally,

$$\Gamma[\bar{B}\to (D\pi)_{\rm res}+\ell\bar{\nu}]=\,\left({2\over \pi}\tan
^{-1}2N\right){1\over \Gamma_{D^*}}\cdot{F(m^2)\over m}+A,\eqno(B.12a)$$
$$\Gamma[\bar{B}\to (D\pi)_{\rm nonres}+\ell\bar{\nu}]=\,\left(1-{2\over \pi}
\tan^{-1}2N\right){1\over \Gamma_{D^*}}\cdot{F(m^2)\over m}+A'.\eqno(B.12b)$$

\noindent
The constant terms $A$ and $A'$ which are independent of $\Gamma_{D^*}$ may
arise from the nonleading contributions of $\Gamma_{\rm D^*\,pole}$, the
$B^*$-pole contributions, and the interference terms between the $B^*$-pole
and $D^*$-pole contributions.  These constants $A$ and $A'$ are generally
very different for resonant and nonresonant contributions.

   The definition (5.6) for resonant contributions corresponds to $N=3$ and
for this choice we have

$$\Gamma[\bar{B}\to (D\pi)_{\rm res}+\ell\bar{\nu}]=\,0.895\,{1\over
\Gamma_{D^*}}\cdot{F(m^2)\over m}+A,\eqno(B.13a)$$
$$\Gamma[\bar{B}\to (D\pi)_{\rm nonres}+\ell\bar{\nu}]=\,0.105\,
{1\over \Gamma_{D^*}}\cdot{F(m^2)\over m}+A'.\eqno(B.13b)$$

\noindent These linear relations are confirmed by the numerical results
shown in Figs. 7-10 for
$\bar{B}^0 \to D^+\pi^0e^-\bar{\nu}_e$, $\bar{B}^0 \to D^0\pi^+e^-\bar{\nu}_e$,
and $B^- \to D^0\pi^0e^-\bar{\nu}_e$.
The ratio of slopes of the two linear relations is

$${ {\rm (slope)}_{\rm res}\over {\rm (slope)}_{\rm nonres}}=\,{0.895\over
0.105}=\,8.52~~,\eqno(B.14)$$

\noindent which agrees with the slopes in Figs. 7-10, as it can be easily
verified.  The decays $B^- \to D^+\pi^-e^-\bar{\nu}_e$ deserve special
attention.  First of all,
$B^- \to (D^+\pi^-)_{\rm res}e^-\bar{\nu}_e$ is kinematically forbidden
if $\Gamma_{D^{*0}} \le 0.4$ MeV,
since $m(D^+\pi^-)$ always falls outside the resonant condition (5.6).
For $\Gamma_{D^{*0}}$ between 0.4 and 1 MeV, the phase space is so
restricted that the decay rates for
$B^- \to (D^+\pi^-)_{\rm res}e^-\bar{\nu}_e$ are completely negligible.
Furthermore, the decay rate for
$B^- \to (D^+\pi^-)_{\rm nonres}e^-\bar{\nu}_e$ is found to be rather
small and independent of $\Gamma_{D^*}$ as seen in Fig. 10.  This
behavior can be understood by the following considerations.  We notice that
the leading term in the approximation (B.8) vanishes since there is no
phase space for the decay $D^{\ast0} \to D^+\pi^-$ (see the mass values
given in (5.1)).  In addition, from (5.1) we have
$$
  P^2 = (p + q)^2 \ge (m_{D^+} + m_{\pi^-})^2, \eqno(B.15)
$$
or
$$
  P^2 - m_{D^{\ast0}}^2 \ge 7500~{\rm MeV}^2, \eqno(B.16)
$$
as compared with
$$
  200.7~{\rm MeV}^2 \le m_{D^{\ast0}} \Gamma_{D^{*0}}
   \le 2007~{\rm MeV}^2, \eqno(B.17)
$$
for $0.1~{\rm MeV} \le \Gamma_{D^{*0}} \le 1~{\rm MeV}$.  We conclude
that in the denominator of the $D^\ast$ propagator of (B.1), the
imaginary part is always small and hence it can be neglected.  The result
is therefore independent of $\Gamma_{D^{*0}}$.  Moreover, the denominator
of the $D^\ast$ propagator is never very small and the phase space near
its minimum (B.15) is very limited.  As a consequence, the decay rate
is substantially reduced.

   The definition (5.6) for resonant contributions is reasonable, but somewhat
arbitrary. However, the decay rate for resonant contributions is rather
insensitive to the definition. As $N$ varies from $N=2$ to $N=\infty$, the
slope in (B.13a) changes by $-5.7\%$ and $+11.7\%$, respectively. On the
other hand, the nonresonant contributions are very sensitive to the
experimental cuts.

   Finally, Figs. 7-10 show that the straight lines for the resonant
contributions for both  charged and neutral $\bar{B}$ mesons pass through the
origin. We conclude that $A\approx 0$. This is to be expected since the cut
(5.6) gives rise to a very small phase space contributing to the constant
$A$.

\nicebreak

\centerline{\bf REFERENCES}

\bigskip

\begin{enumerate}

\item N. Isgur and M.B. Wise, {\em Phys. Lett.} {\bf B232}, 113 (1989);
{\em Phys. Lett.} {\bf B237}, 527 (1990).

\item M.B. Voloshin and M.A. Shifman, {\em Yad. Fiz.} {\bf 45}, 463 (1987)
[{\em Sov. J. Nucl. Phys.} {\bf 45}, 292 (1987)].

\item T.M. Yan, H.Y. Cheng, C.Y. Cheung, G.L. Lin, Y.C. Lin, and H.L. Yu,
{\em Phys. Rev.} {\bf D46}, 1148 (1992).  See also T.M. Yan, {\em Chin. J.
Phys.} (Taipei) {\bf 30}, 509 (1992).

\item M.B. Wise, {\em Phys. Rev.} {\bf D45}, R2188 (1992).

\item G. Burdman and J. Donoghue, {\em Phys. Lett.} {\bf B280}, 287 (1992).

\item P. Cho, {\em Phys. Lett.} {\bf B285}, 145 (1992).

\item Particle Data Group, {\em Phys. Rev.} {\bf D45}, S1 (1992).

\item The ARGUS Collaboration, H. Albrecht {\it et al.}, {\em Z. Phys.}
{\bf C57}, 533 (1993).

\item The CLEO Collaboration, R. Fulton {\it et al.}, {\em Phys. Rev.} {\bf
43}, 651 (1992); The CLEO Collaboration, S. Hernandez {\it et al.}, ibid
{\bf D45}, 2212 (1992).

\item Clarence L.Y. Lee, Ming Lu, and Mark B. Wise, {\em Phys. Rev.} {\bf
D46}, 5040 (1992).

\item Clarence L.Y. Lee, CALT-68-1841 (1992).

\item G. Kramer and W.F. Palmer, DESY 92-130 (1992);
G. K\"opp, G. Kramer, W. Palmer, and G. Schuler, {\em Z. Phys.} {\bf
C48}, 327 (1990).

\item A. Pais and S.B. Treiman, {\em Phys. Rev.} {\bf 168}, 1858 (1968);
see also, N. Cabibbo and A. Maksymowicz, {\em Phys. Rev.} {\bf 137}, B438
(1965).

\item H.Y. Cheng, C.Y. Cheung, G.L. Lin, Y.C. Lin, T.M. Yan, and H.L. Yu,
{\em Phys. Rev.} {\bf D47}, 1030 (1993).

\item The CLEO Collaboration, F. Butler {\it et al.}, {\sl Phys. Rev. Lett.}
{\bf 69}, 2041 (1992).

\item G.L. Lin, invited talk delivered at the First Workshop on Particle
Physics Phenomenology, Kenting, Taiwan, R.O.C., May 1992, IP-ASTP-16-92,
to appear in Chin. J. Phys. (1993); H.Y. Cheng, IP-ASTP-18-92, to appear in
{\it Proceedings of the XXVI International Conference on High Energy Physics},
Dallas, August 1992.

\item G.L.Kane, K. Stowe, and W.B. Rolnick, {\em Nucl. Phys.} {\bf B152},
390 (1979).

\item P. Nyborg and O. Skjeggestad, in {\em Kinematics and Multiparticle
Systems}, Proceedings of the International School of Elementary Particle
Physics, Herceg-Novi, Yugoslavia, 1965, edited by M. Nikoli\'c.
Pages 55-57 are especially useful for computing single particle energy spectra.

\item N. Cabibbo, {\em Phys. Rev. Lett.} {\bf 10}, 531 (1963); M. Kobayashi
and K. Maskawa, {\em Prog. Theor. Phys.} {\bf 49}, 652 (1993).

\item The value of the $B^*$ mass used here is from the 1990 edition of PDG
[{\sl Phys. Lett.} {\bf B239}, S1 (1990)]. We have checked our results using
its new value in 1992 PDG, $m_{B^*}=5324.6$ MeV.
The differences for $\bar{B}\to D\pi\ell\bar{\nu}$ are negligible and those for
$\bar{B}\to D^*\pi\ell\bar{\nu}$ are less than $1\%$.

\item G. Burdman, {\em Phys. Lett.} {\bf B284}, 133 (1992).

\item For a review of the experimental status of $V_{cb}$, see Persis S. Drell
and J. Richie Patterson, CLNS 92/1177, to appear in
{\it Proceedings of the XXVI International Conference on High Energy Physics},
Dallas, August 1992.

\item We have benefited from conversations with Prof. P. Drell on the general
question of how to define resonant and nonresonant contributions.

\item P. Bialas, K. Zalewski, and J.G. K\"orner, TPJU-19/92, MZ-TH/92-53
(1992); Ming Lu, M.B. Wise, and N. Isgur, {\sl Phys. Rev.} {\bf D45}, 1553
(1992).

\end{enumerate}

\nicebreak

\centerline{\bf Figure Captions}

\bigskip

\begin{description}

\item[Fig. 1] Feynman diagrams for $\bar{B}\to D\pi\ell\bar{\nu}$.

\item[Fig. 2] Feynman diagrams for $\bar{B}\to D^\ast\pi\ell\bar{\nu}$.

\item[Fig. 3] {\bf (a)} General kinematics of
$\bar{B}\to D(D^\ast) \pi\ell\bar{\nu}$.
The dashed lines are the lines of flight of the $D(D^\ast) \pi$ pair and
lepton pair in the rest frame of the $\bar{B}$ meson.  Solid lines denote the
line of flight of $D$ $(D^\ast)$ and $\pi$ in the $M$-frame and that of
the lepton and neutrino in the $L$-frame.  The coordinates $x, ~y$, and $z$
are indicated, and the angles $\theta$, $\theta_\ell$ and $\phi$ are
labeled.
{\bf (b)} The three linear polarization vectors of $D^\ast$ in the
$M$-frame.  The vector ${\vec{e}}_1$ is in x-z plane, ${\vec{e}}_2$ is
along the ${\hat{y}}$ axis and ${\vec{e}}_3$ is along ${\vec{p}}$.

\item[Fig. 4] The energy spectra of the $D$ meson from
$\bar{B}^0 \to D^+\pi^0e^-\bar{\nu}_e$ in the resonant region
(solid line) and in the nonresonant region (broken lines).  Shown
are effects of two pion momentum cutoffs (100 MeV and 200 MeV) in
the $D$-frame on the nonresonant contributions.  The bump at the high energy
end is an artifact of the simple cutoff imposed on the pion.  The resonant
contribution is not affected by these pion momentum cutoffs.  The spectrum is
in units of MeV / MeV.

\item[Fig. 5] The energy spectra of the electron from
$\bar{B}^0 \to D^+\pi^0e^-\bar{\nu}_e$ in the resonant region
(solid line) and in the nonresonant region (broken lines).  Shown
are effects of two pion momentum cutoffs (100 MeV and 200 MeV) in
the $D$-frame on the nonresonant contributions.  The resonant
contribution is not affected by these pion momentum cutoffs.  The spectrum is
in units of MeV / MeV.

\item[Fig. 6] The energy spectra of the pion from
$\bar{B}^0 \to D^+\pi^0e^-\bar{\nu}_e$ in the resonant region
(solid line) and in the nonresonant region (broken lines).  Shown
are effects of two pion momentum cutoffs (100 MeV and 200 MeV) in
the $D$-frame on the nonresonant contributions.  The resonant
contribution is not affected by these pion momentum cutoffs.  The spectrum is
in units of MeV / MeV.

\item[Fig. 7] The decay rates
$\Gamma(\bar{B}^0 \to D^+\pi^0e^-\bar{\nu}_e)$ (labeled as
neutral pion) and $\Gamma({\bar{B}}^0 \to D^0\pi^+e^-\bar{\nu}_e)$
(labeled as charged pion) in the resonant region as a function of
$1/\Gamma_{D^*}$.  The pion momentum cutoff of 100 MeV or 200 MeV in the
$D$-frame has no effect on these rates.

\item[Fig. 8] The decay rates
$\Gamma({\bar{B}}^0 \to D^+\pi^0e^-\bar{\nu}_e)$ (labeled as
neutral pion) and $\Gamma({\bar{B}}^0 \to D^0\pi^+e^-\bar{\nu}_e)$
(labeled as charged pion) in the nonresonant region as a function of
$1/\Gamma_{D^*}$.  The pion momentum cutoff is 100 MeV in the $D$-frame.

\item[Fig. 9] The decay rates
$\Gamma(B^- \to D^0\pi^0e^-\bar{\nu}_e)$ (labeled as
neutral pion) in the resonant region as a function of
$1/\Gamma_{D^*}$.  The other mode
$B^- \to D^+\pi^-e^-\bar{\nu}_e$ is kinematically forbidden
in the resonant region.
The pion momentum cutoff of 100 MeV or 200 MeV in the $D$-frame has no effect
on these rates.

\item[Fig. 10] The decay rates
$\Gamma(B^- \to D^0\pi^0e^-\bar{\nu}_e)$ (labeled as
neutral pion) and $\Gamma(B^- \to D^+\pi^-e^-\bar{\nu}_e)$
(labeled as charged pion) in the nonresonant region as a function of
$1/\Gamma_{D^*}$.  The pion momentum cutoff is 100 MeV in the $D$-frame.
See Appendix B for an explanation of the different behavior of the two
decay rates.

\item[Fig. 11] The energy spectra of the $D^\ast$ meson with different
polarizations in the $B$-frame from
$\bar{B}^0 \to D^{\ast+}\pi^0e^-\bar{\nu}_e$
with a pion momentum cutoff of 100 MeV in the $B$-frame.

\item[Fig. 12] The energy spectra of the electron from
$\bar{B}^0 \to D^{\ast+}\pi^0e^-\bar{\nu}_e$ for different $D^\ast$
polarizations in the $B$-frame with a pion momentum cutoff of 100 MeV
in the $B$-frame.

\item[Fig. 13] The energy spectra of the pion from
$\bar{B}^0 \to D^{\ast+}\pi^0e^-\bar{\nu}_e$ for different $D^\ast$
polarizations in the $B$-frame with a pion momentum cutoff of 100 MeV
in the $B$-frame.

\item[Fig. 14] The energy spectra of the $D^\ast$ meson with different
polarizations in the $B$-frame from
$\bar{B}^0 \to D^{\ast+}\pi^0e^-\bar{\nu}_e$
with a pion momentum cutoff of 100 MeV in both the $B$- and $D^\ast$-frames.

\item[Fig. 15] The energy spectra of the electron from
$\bar{B}^0 \to D^{\ast+}\pi^0e^-\bar{\nu}_e$ for different $D^\ast$
polarizations in the $B$-frame with a pion momentum cutoff of 100 MeV
in both the $B$- and $D^\ast$-frames.

\item[Fig. 16] The energy spectra of the pion from
$\bar{B}^0 \to D^{\ast+}\pi^0e^-\bar{\nu}_e$ for different $D^\ast$
polarizations in the $B$-frame with a pion momentum cutoff of 100 MeV
in both the $B$- and $D^\ast$-frames.

\item[Fig. 17] The ratio $d \Gamma ({\bar{B}}^0 \to D^{\ast +}
\pi^0 e^- {\bar{\nu}})/d \Gamma ({\bar{B}}^0 \to D^+ \pi^0 e^-
{\bar{\nu}})$
for different $D^\ast$ polarizations in the $B$-frame
as a function of $v \cdot v^\prime$. The $D^+ \pi^0$ system
is in the resonant region.  The pion momentum cutoff is 100 MeV in both the
$B$- and $D^\ast (D)$-frame.

\item[Fig. 18] The ratio $d \Gamma ({\bar{B}}^0 \to D^{\ast +}
\pi^0 e^- {\bar{\nu}})/d \Gamma ({\bar{B}}^0 \to D^+ \pi^0 e^-
{\bar{\nu}})$
for different $D^\ast$ polarizations in the $B$-frame
as a function of $v \cdot v^\prime$. The $D^+ \pi^0$ system
is in the nonresonant region.  The pion momentum cutoff is 100 MeV in both
the $B$- and $D^\ast (D)$-frame.

\item[Fig. 19] The ratio $d \Gamma ({\bar{B}}^0 \to D^{\ast +}
\pi^0 e^- {\bar{\nu}})/d \Gamma ({\bar{B}}^0 \to D^+ \pi^0 e^-
{\bar{\nu}})$
for different $D^\ast$ polarizations in the $B$-frame
as a function of $v \cdot v^\prime$. The $D^+ \pi^0$ system
is in the resonant region.  The pion momentum cutoff is 200 MeV in both
the $B$- and $D^\ast (D)$-frame.

\item[Fig. 20] The ratio $d \Gamma ({\bar{B}}^0 \to D^{\ast +}
\pi^0 e^- {\bar{\nu}})/d \Gamma ({\bar{B}}^0 \to D^+ \pi^0 e^-
{\bar{\nu}})$
for different $D^\ast$ polarizations in the $B$-frame
as a function of $v \cdot v^\prime$. The $D^+ \pi^0$ system
is in the nonresonant region.  The pion momentum cutoff is 200 MeV in both
the $B$- and $D^\ast (D)$-frame.

\item[Fig. 21] The decay rates $ \Gamma ({\bar{B}}^0 \to D^+
\pi^0 e^- {\bar{\nu}})$ for nonresonant $D^+ \pi^0$ as a function of the
pion momentum cutoff in the $D$-frame.

\item[Fig. 22] The decay rates $ \Gamma ({\bar{B}}^0 \to D^{\ast +}
\pi^0 e^- {\bar{\nu}})$ as a function of the pion momentum cutoff in the
$B$-frame.

\item[Fig. 23] The decay rates $ \Gamma ({\bar{B}}^0 \to D^{\ast +}
\pi^0 e^- {\bar{\nu}})$ as a function of the pion momentum cutoff in both the
$B$- and $D^\ast$-frame.

\end{description}

\end{document}